\documentclass[conference,letterpaper,11pt]{IEEEtran}

\usepackage{todonotes}
\usepackage[utf8]{inputenc} 
\usepackage[T1]{fontenc}
\usepackage{url}
\usepackage{ifthen}
\usepackage{cite}
\usepackage[cmex10]{amsmath} 

\usepackage[english]{babel}
\usepackage{amssymb}

\usepackage{enumitem}
\usepackage{empheq}
\usepackage{graphicx}

\def\F{\mathbb{F}}

\newtheorem{lem}{Lemma}
\newtheorem{thm}{Theorem}
\newtheorem{cor}{Corollary}
\newtheorem{defin}{Definition}
\newtheorem{remark}{Remark}

\newcommand{\Mod}[1]{\ (\mathrm{mod}\ #1)}

\def\aL{\alpha_{\mathrm{lin}}}

\def\th{\vartheta}

\def\o{\overline}
\def\tl{\rho_{\mathrm{lin}}}

\begin{document}
\title{Linear Shannon Capacity of Cayley Graphs} 

\author{%
  \IEEEauthorblockN{Venkatesan Guruswami and Andrii Riazanov}
  \IEEEauthorblockA{Carnegie Mellon University\\
                    Computer Science Department\\
                    Pittsburgh, PA 15213\\
                    Email: \{venkatg,\ riazanov\}@cs.cmu.edu}
}

\maketitle

\begin{abstract}
The Shannon capacity of a graph is a fundamental quantity in zero-error information theory measuring the rate of growth of independent sets in graph powers. Despite being well-studied, this quantity continues to hold several mysteries. Lov\'asz famously proved that the Shannon capacity of $C_5$ (the 5-cycle) is at most $\sqrt{5}$ via his theta function. This bound is achieved by a simple linear code over $\F_5$ mapping $x \mapsto 2x$. 

\smallskip
This motivates the notion of \emph{linear Shannon capacity} of graphs, which is the largest rate achievable when restricting oneself to linear codes.  We give a simple proof based on the polynomial method that the linear Shannon capacity of $C_5$ is $\sqrt{5}$. Our method applies more generally to Cayley graphs over the additive group of finite fields~$\F_q$, giving an upper bound on the linear Shannon capacity.
We compare this bound to the  Lov\'asz theta function, showing that they match for self-complementary Cayley graphs (such as $C_5$), and that the bound is smaller in some cases. We also exhibit a quadratic gap between linear and general Shannon capacity for some graphs.
\end{abstract}

\section{Introduction}

For graphs $G_1, G_2, \dots, G_k$, the strong product $G_1\boxtimes G_2 \boxtimes \dots \boxtimes G_k$ is a graph with vertex  set $V(G_1) \times V(G_2) \times \dots \times V(G_k)$, where $(v_1, v_2, \dots, v_k) \sim (u_1, u_2, \dots, u_k)$ if and only if $v_i = u_i$ or $(v_i, u_i) \in E(G_i)$ for every $i \in \{1, 2, \dots, k\}$. For the strong product of copies of the same graph we write $G^k = G\boxtimes G \boxtimes \dots \boxtimes G$. 

The Shannon capacity of a graph $G$, introduced by Shannon in~\cite{Shannon}, is defined as 
\[ \Theta(G) = \sup\limits_k \sqrt[k]{\alpha\left(G^k\right)} = \lim\limits_{k \to \infty}\sqrt[k]{\alpha\left(G^k\right)}  \ , \]
where $\alpha(\cdot)$ denotes the independence number of the graph.
Determining the Shannon capacity of an arbitrary graph is a very difficult problem, and very little is known about it in general, despite a considerable amount of attention it has received in the information theory and combinatorics communities (\cite{Lovasz, Haemers, Alon_shan, Schrijver, Bohman}). In particular, $\Theta(C_7)$ is still unknown, where $C_7$ is the $7$-cycle. For $C_5$, the famous work of Lov\'{a}sz that introduced the theta function proved that the Shannon capacity equals~$\sqrt{5}$~\cite{Lovasz}.

 In coding theory, $\Theta(G)$ captures the zero-error capacity of the channel with confusion graph $G$. Specifically, consider a coding channel with input set $V = \{1, 2, \dots, n\}$, and let the confusion graph $G$ have $V$ as the vertex set. Further, let $(v, u) \in E(G)$ if and only if the letters $v$ and $u$ might be confused in the transmission (i.e. lead to the same output). Clearly, $\alpha(G)$ captures the maximum size of a set of letters that can be communicated in an error-free manner in a single use of the channel. From the definition of the graph power, it follows that $\alpha(G^k)$ represents the largest set of $k$-letter words (code) that can be communicated in an error-free manner over $k$ uses of the channel. Therefore, $\Theta(G)$ can be interpreted as the maximal effective number of symbols that can be transmitted per use of the channel, amortized over $k$ uses of the channel in the limit of large~$k$.

In this paper, we consider a special case of zero-error communication described above, where we restrict the codes to be linear. Linear codes appear in various contexts of coding theory, and they often can match the guarantees of general codes, especially in their limiting behaviour. This restriction is thus well-motivated so one can understand how well \emph{linear} codes allow communication in this zero-error regime, and whether they might be able to achieve or come close to the Shannon capacity in interesting cases (of course, when restricting the graph to have a prime power number of vertices). For instance, we know that for $C_5$, the Shannon capacity is achieved by a linear code $\{(x, 2x) \mid x\in \F_5\}$. More generally, linear codes can achieve the Shannon capacity for any Paley graph (a graph with vertices in a finite field $\F_q$ with $q \equiv 1\, \Mod{4}$, where $u, v \in \F_q$ are adjacent if and only if $(u - v)$ is a quadratic residue in $\F_q$). On the flip side, it is interesting to investigate whether one can prove better upper bounds on the Shannon capacity when restricting to linear codes, or prove similar bounds in a simpler manner.

The main result in this paper (Theorem~\ref{theorem_main}) is an upper bound on the zero-error capacity that can be achieved using only linear codes, for the case when $G = \Gamma(\F_q, S)$ is a Cayley graph over the additive group of a finite field $\F_q$ with a symmetric set $S$. The upper bound is proven by an application of the polynomial method. 

The same upper bound appeared previously in the literature in~\cite{Sperner}, as was brought to authors' attention during the review process. While our result itself is thus not new, our proof is different and independent of the previous work. We also hope that our will renew attention on the interesting concept of linear Shannon capacity.

We also compare the above upper bound to the Lov\'asz theta function $\th(G)$, which is a general upper bound on Shannon capacity formulated as a solution to a semidefinite program~\cite{Lovasz}. We show that for any $G$ as above, this upper bound either coincides with the theta function for both $G$ and the complement graph $\o{G}$, or it is strictly stronger for one of $G$ and $\o{G}$. We then show an example of graphs for which there is a quadratic separation between our upper bound and the true value of the Shannon capacity. This 
shows a gap between the performance of linear and general codes in this context.


On the other hand, the upper bound coincides with the Lov\'asz theta function for any Paley graph ($C_5$ in particular), and more generally for any \emph{self-complementary} Cayley graph as above. Even though such a bound also follows from~\cite{Lovasz}, this gives an alternative simple proof, using the polynomial method, of the best possible rates achievable by linear codes for such graphs.

One interesting direction for future investigation would be to prove better upper bounds on the linear Shannon capacity for odd cycles $C_p$ for prime $p$. (The upper bound in Theorem~\ref{theorem_main} is worse than the Lov\'asz theta function for $C_p$ for $p > 5$, though it is stronger for the complement $\o{C_p}$.) In particular, the case $p = 2^k + 1$ is intriguing, because (to the best of our knowledge) the best lower bound on Shannon capacity $\Theta(C_p) \geq p^{(k-1)/k}$ for this case actually comes from a linear code construction from~\cite{baumert}.  

Another interesting question is whether one can somehow argue that the best codes for $C_5$ must be affine. Such a reduction together with the bound from Theorem~\ref{theorem_main} would then suffice to pin down the Shannon capacity of $C_5$ without resorting to semidefinite programming. More broadly, we hope that the twist of considering the power and limitations of linear codes for zero-error communication might provide a fruitful new perspective in the study of the classic Shannon capacity problem and its variants.

\section{Upper bound on the linear Shannon capacity}
Since we are interested in linear codes, we restrict the vertex set of $G$ to be a finite field $\F_q$, where $q$ is a prime power. 
We first define linear independence number for powers of $G$:
\begin{defin}[Implicit in \cite{Sperner}]
For a graph $G$ with $V(G) = \F_q$ and any $k \geq 1$, the \emph{linear independence number} of $G^k$, denoted as $\aL(G^k)$, is the size of the largest independent set $I_L$ of $G^k$ that is linear, i.e. which can be represented as $I_L = \{ (x, Ax), \text{\,for } x\in \F^m_q\}$ for some matrix $A \in \F_q^{(k-m)\times m}$, where $1 \leq m \leq k$. Equivalently, $\aL(G^k)$ is the size of the largest independent set of $G^k$ that is a linear subspace of $\F_q^k$.
\end{defin}

As with the independence number, it is easy to see that $\aL(G^{k+d}) \geq \aL(G^k)\cdot\aL(G^d)$. Indeed, if $I_L^{(k)} = \{ (x, A^{(k)}x), \text{\,for } x\in \F^{m_k}_q\}$ and $I_L^{(d)} = \{ (x, A^{(d)}x), \text{\,for } x\in \F^{m_d}_q\}$ are the largest linear independent sets of $G^k$ and $G^d$, then the set $I_L^{(k)} \times I_L^{(d)}$ is also independent and linear, since $I_L^{(k)} \times I_L^{(d)} = \{ (x, y,  A^{(k)}x, A^{(d)}y), \text{\,for } (x, y)\in \F^{m_k + m_d}_q\}$. 

We then define the linear Shannon capacity of $G$ as \[\Theta_{\mathrm{lin}}(G) \coloneqq \sup\limits_k\sqrt[k]{\aL\left(G^k\right)} = \lim\limits_{k \to \infty}\sqrt[k]{\aL\left(G^k\right)} \ ,\] 
where the equality follows from the above supermultiplicativity and Fekete's lemma (same as for $\Theta(G)$). 

Let $\F_q$ be a finite field, and take a set $S \subseteq \F_q$ which is symmetric under addition, i.e. $S = -S$, and $0 \notin S$. We then consider Cayley graph $G = \Gamma(\F_q, S)$, which has vertex set $V(G) = \F_q$ and for which $(v, u) \in G$ if and only if $(v - u) \in S$. Notice that we don't require $S$ to be a generating set of $\F_q$, as is the usual definition of Cayley graphs.

We now prove our main result, which is an upper bound on the linear Shannon capacity for such Cayley graphs. As mentioned in the introduction, the same bound appeared in the literature in~\cite[Corollary 2.1]{Sperner}, where it was derived using Jamison's theorem from~\cite{Jamison}.

\begin{thm}[Main]
\label{theorem_main}
Let $\F_q$ be any finite field, and $S\subseteq \F_q \setminus \{0\}$ be any symmetric set. Then
\[ \Theta_{\mathrm{lin}}\big(\Gamma(\F_q, S)\big) \leq q^{1 - \frac{|S|}{q-1}}. \]

\end{thm}
\begin{IEEEproof} Denote $s = |S|$, $G = \Gamma(\F_q, S)$, fix any $n > 1$, and let $I_L^{(n)}$ be the largest linear independent set of $G^n$. Let this set be  $I_L^{(n)} = \{ (x, Ax), \text{\,for } x\in \F^m_q\}$ for some matrix $A \in \F_q^{(n-m)\times m}$. Since $I_L^{(n)}$ is an independent set, there is no $x, y \in \F_q^m$ such that $(x, Ax) \sim (y, Ay)$ in~$G^n$, where ``$\sim$'' denoted adjacency. Equivalently, there is no $z \in \F_q^m$ such that $(z, Az) \sim 0^n$ in $G^n$, due to the linearity of the code and the additive structure of Cayley graphs.

Denote further $S_0 = S \cup \{0\}$. 
It is clear that $z \sim 0^m$ if and only if $z \in (S_0)^m \setminus \{0^m\}$. Therefore,  since $(z, Az) \nsim 0^n$ for all $z \in \F_q^m$, it follows that $Az \nsim 0^{(n-m)}$ for all $z \in (S_0)^m \setminus \{0^m\}$, or, in other words, $Az \notin (S_0)^{(n-m)}$ for all such $z$. Denote by $D$ the complement of $S_0$: $D = \F_q \setminus S_0$. It then follows that for any $z \in (S_0)^m \setminus \{0^m\}$ there exists a coordinate $i$ such that $(Az)_i \in D$. Define the following polynomial in $\F_q[z_1, z_2, \dots, z_m]$:
\[ P(z) = \prod_{i=1}^{n-m} \prod_{d \in D}\Big( \langle a_i, z\rangle - d\Big),\]
where $a_i$ is the $i^{\text{th}}$ row of $A$. 

From the argument above it follows that $P(z) = 0$ for all $z \in (S_0)^m \setminus \{0^m\}$. We will also denote the constant $c = P(0) = \left(\prod\limits_{d\in D} d\right)^{n-m}$. Note that 
\[ \deg(P) \leq (n-m)\cdot (q-1-s) \ .\]

Consider now the ideal $R$ of $\F_q[z_1, z_2, \dots, z_m]$, generated by the polynomials $\prod\limits_{t \in S_0} (z_i-t)$, for all $i = 1, 2, \dots, m$. Let further $Q(z)$ be a remainder of $P(z)$ modulo the ideal $R$. Since every polynomial in $R$ takes value $0$ for any $z \in (S_0)^m$ by construction, it follows that $Q(z) = P(z)$ for all $z \in (S_0)^m$. Moreover, it is clear that $\deg_{z_i}(Q) \leq s$ for all $i = 1, 2, \dots, m$, since $R$ contains degree-$(s+1)$ univariate polynomials in every variable $z_i$.

We now use the following well-known lemma, which appears in the literature in the context of low-degree testing, and is a special case of the Combinatorial Nullstellensatz~\cite{alon}.

\begin{lem}[Low Degree Extension]
Let $\F$ be any field, and let $H\subset \F$ be a set of size $h$. 
Then any function $f : H^m \to \F$ can be \em uniquely \em extended to an $m$-variate polynomial $G : \F^m \to \F$, such that $G$ has degree at most $h-1$ in each variable. 
\end{lem}

Taking in our settings $H = S_0 \subset \F_q$, the above lemma implies that there exists a \emph{unique} polynomial $G(z)$ such that $G(0^m) = c$ and $G(z) = 0$ for all $z\in \left(S_0\right)^m \setminus \{0^m\}$, and for which $\deg_{z_i}(G) \leq s$ for all $i=1, 2, \dots, m$. Now notice that the polynomial $G(z) = c_2\cdot \prod\limits_{i = 1}^m \prod\limits_{t\in S}(z_i - t)$ meets all these conditions (where $c_2 \in \F_q$ is a normalizing constant which ensures that $G(0^m) = c = P(0))$, and the same holds for $Q(z)$ by construction, as discussed above. Therefore, out of uniqueness, we conclude 
\[ Q(z) = c_2\cdot \prod\limits_{i = 1}^m \prod\limits_{t\in S}(z_i - t) \, \] 
and in particular it means $\deg(Q) = sm$.

Finally, we derive $sm = \deg(Q) \leq \deg(P) \leq (n-m)\cdot(q-1-s)$, and therefore we obtain the bound on the rate of the linear code $I_L^{(n)}$:
\[ \dfrac{m}{n} \leq 1 - \dfrac{s}{q-1}.\]
Since $\aL(G^n) = I_L^{(n)} = q^m$, this precisely gives the desired bound $\sqrt[n]{\aL(G^n)} \leq  q^{1 - \frac{s}{q-1}}$ for any $n > 1$, and so the bound holds for $\Theta_{\mathrm{lin}}(G)$, as stated.
\end{IEEEproof}
\begin{remark}
The bound in Theorem~\ref{theorem_main} only depends on~$|S|$, and the proof relies only on degree-based arguments. An interesting question for further investigation is to improve the bounds on $\Theta_{\mathrm{lin}}(G)$ by using the product structure of the polynomials $P(z)$ and $Q(z)$, possibly for the case when some additional structure of the set $S$ is given.
\end{remark}

\section{Comparison to Shannon capacity and Lov\'asz theta function}

Let $G = \Gamma(\F_q, S)$ be a Cayley graph as in Theorem~\ref{theorem_main}. It is straightforward to see that the complement of $G$ is $\overline{G} = \overline{\Gamma(\F_q, S)} = \Gamma(\F_q, \o{S})$, where we denote $\o{S} = \left(\F_q \setminus \{0\}\right)\setminus S$. So $\o{G}$ is also a Cayley graph for which Theorem~\ref{theorem_main} applies.

Let $\th(G)$ be the Lov\'asz theta function of the graph $G$, the general upper bound on the (actual) Shannon capacity $\Theta(G)$. It is well known that any Cayley graph is vertex-transitive, therefore it follows from~\cite{Lovasz} that ${\th(G) \cdot \th(\overline{G}) = q}$.




For convenience, we denote the upper bound from Theorem~\ref{theorem_main} as $\rho_{\mathrm{lin}}(G) = q^{1 - \frac{|S|}{q-1}}$. Applying it to $G$ and $\o{G}$ we then obtain
\begin{align*}
\Theta_{\mathrm{lin}}(G)\cdot \Theta_{\mathrm{lin}}(\o{G}) &\leq \tl(G) \cdot \tl(\o{G})  \\ & =  q^{1 - \frac{|S|}{q-1}} \cdot q^{1 - \frac{(q-1) - |S|}{q-1}} = q.
\end{align*}
Since $\th(G) \cdot \th(\overline{G}) = \tl(G) \cdot \tl(\o{G}) = q$, we conclude with the following
\begin{cor}
\label{cor:product}
Let $\F_q$ be any finite field, $S\subseteq \F_q \setminus \{0\}$ be any symmetric set, and let $G$ be the Cayley graph $\Gamma(\F_q, S)$. Then one of the following holds:
\begin{enumerate}[label = (\alph*)]
\item  upper bounds on the linear Shannon capacity from Theorem~\ref{theorem_main} coincide with Lov\'asz theta function for both $G$ and $\o{G}$, e.g. $\tl(G) = \th(G)$ and $\tl(\o{G}) = \th(\o{G})$
\item for one of the graphs $G$ and $\o{G}$, the bound $\tl(\cdot)$ on the linear Shannon capacity is strictly smaller than Lov\'asz theta function $\th(\cdot)$. 
\end{enumerate}
\end{cor}

In the general case, $\tl(G)$ and $\th(G)$ rarely coincide. This can be easily seen by noticing that modifying the set $S$ slightly (while keeping its size the same) doesn't change the upper bound $\tl(G)$, while $\th(G)$ is very likely to change. Therefore, one expects that for typical graphs, the upper bound $\tl(\cdot)$ on the linear Shannon capacity is usually strictly smaller than Lov\'asz theta function for either $G$ or $\o{G}$.



\subsection{Self-complementary graphs}

We now describe one class of graphs for which the upper bound from Theorem~\ref{theorem_main} coincides with the Lov\'{a}sz theta function as well as the Shannon capacity. This is the class of self-complementary graphs. For a Cayley graph $G=\Gamma(\F_q,S)$ to be self-complementary, we must have $q \equiv 1 \Mod{4}$, and $|S| = \frac{q-1}{2}$. Then Theorem~\ref{theorem_main} immediately yields $\tl(G) = \tl(\o{G}) = \sqrt{q}$. 

Further, since any such Cayley graph $G$ is self-complementary and vertex-transitive, it is known that $\th(G) = \Theta(G) = \th(\o{G}) = \Theta(\o{G}) = \sqrt{q}$~\cite{Lovasz}. Thus, in this case the upper bound on the linear Shannon capacity from Theorem~\ref{theorem_main}, the Lov\'asz theta function, and the actual Shannon capacity all coincide for both $G$ and $\o{G}$. Though the upper bound of $\sqrt{q}$ on linear Shannon capacity follows from~\cite{Lovasz}, our approach gives an alternative proof of this bound (for linear codes only), without appealing to the Lov\'asz theta function, or semidefinite programming in general.

The next natural question we ask in this context is when the bound $\Theta_{\mathrm{lin}}(G) \leq \sqrt{q}$ on linear Shannon capacity is tight and achieved with equality. When this is the case, it means that linear codes can exactly match the performance of general codes for zero-error communication over the corresponding graph. Below we discuss two families of graphs for which the equality is achieved.

For both examples below where we show that $\Theta_{\mathrm{lin}}(G) = \sqrt{q}$, this equality is achieved by having a linear independent set (of dimension $1$, or size $q$) in the second power of the graph $G$. This just means that $I_a = \{(x, ax)\,|\, x \in \F_q \}$ is an independent set in $G^2$ for some $a \in \F_q$. Since $(0, 0) \in I_a$ and out of linearity (additive symmetry) of the code, this happens if and only if $0 \sim x \Leftrightarrow 0 \not\sim ax$. This is then equivalent to saying that $x \mapsto ax$ is a \emph{linear complementing isomorphism} of $G$, i.e. this is a multiplicative map which is an isomorphism from $G$ to its complement $\overline{G}$.

The first case when any self-complementary graph has such a linear complementing isomorphism is when $q = p$ is a prime number. Notice that in this case the graph $G$ is circulant (a Cayley graph of a cyclic group). The existence of a linear complementing isomorphism for this case follows directly from a "corrected" Adam's conjecture, initially formulated in~\cite{Adam} and then proven for the prime case in~\cite{Djok}. This conjecture (simplified for our purposes) states that if $\Gamma(\F_p, R)$ and $\Gamma(\F_p, T)$ are isomorphic for some sets $R$ and $T$ (symmetric around $0$), then $R = a\cdot T$ for some $a \in \F_p$. Therefore, if $G = \Gamma(\F_p, S)$ is isomorphic to $\overline{G} = \Gamma(\F_p, \overline{S})$, then $\overline{S} = a\cdot S$, and it clearly follows that $x \mapsto ax$ is a linear complementing isomorphism for $G$. Therefore, $\Theta_{\mathrm{lin}}(G) = \sqrt{p}$ for a prime $p$.

The second case we consider is a well-known family of Paley graphs $P_q$ for any prime power $q$ such that $q\equiv 1 \Mod{4}$, where $P_q = \Gamma(\F_q, S)$ for $S$ being a set of all quadratic residues in $\F_q$. The linear complementing isomorphism which takes $P_q$ to $\overline{P_q}$ is $x \mapsto ax$ for any quadratic non-residue $a \in \F_q$. 

As a special case of the both examples above, one can consider the $5$-cycle $C_5$, which is the Paley graph of order $5$. Therefore, Theorem~\ref{theorem_main} combined with the above discussion gives a simple proof that $\Theta_{\mathrm{lin}}(C_5) = \sqrt{5}$ using the polynomial method.

\subsection{Cayley graphs with quadratic gap between linear and general Shannon capacity}

We conclude the paper with an an example of a family of graphs, for which $\tl(G)$ is strictly smaller than the actual Shannon capacity. This proves a separation between linear and general codes for the zero-error capacity setting.

Consider the case when the finite field is $\F_p$ for a prime $p$, and consider the set 
\[ S = \left\{ \frac{p-1}{4} + 1, \frac{p-1}{4} + 2, \dots, \frac{3(p-1)}{4}\right\} \ . \]
As before, let $G = \Gamma(\F_p, S)$. It is easy to see that $\alpha(G) \geq \frac{p+3}{4}$, since the set $\left\{0, 1, \dots, \frac{p-1}{4}\right\}$ is an independent set in $G$. This immediately gives a lower bound on Shannon capacity, and thus on the Lov\'asz theta function: $\th(G) \geq \Theta(G) \geq  \frac{p+3}{4}$. On the other hand, $\tl(G) = \sqrt{p}$ since $|S| = \frac{p-1}{2}$, and so the linear Shannon capacity $\Theta_{\mathrm{lin}}(G)$ is bounded above by $\sqrt{p}$. This means that there is a quadratic separation between linear and actual Shannon capacity for this family of graphs:
\[ \Theta_{\mathrm{lin}}(G) \leq  \tl(D) = \sqrt{p} < \frac{p+3}{4} \leq \Theta(G) \leq \th(G). \]
One can notice that tweaking the set $S$ slightly will still result in different graphs for which the Shannon capacity is linear in $p$, while the linear Shannon capacity is bounded by $\sqrt{p}$. So such a separation is not specific to just this particular example, but happens for a broader range of graphs.

\newpage
\section*{Acknowledgment}
This work was supported in part by NSF grants CCF-1563742 and CCF-1814603.

We thank the anonymous reviewers who provided valuable comments about the paper and brought the reference~\cite{Sperner} to our attention.




\bibliographystyle{IEEEtran}
\bibliography{bib}









\end{document}